\documentclass[onecollarge,natbib]{svjour2}
\bibpunct{[}{]}{;}{n}{}{,} % to get "[numbered]" references from natbib
\smartqed  % flush right qed marks, e.g. at end of proof
\usepackage{graphicx}
\usepackage{epsfig}
\begin{document}

\title{Hadrons in AdS/QCD models
%\thanks{Presented by T. Frederico at LIGHTCONE 2011, 23 - 27 May, 2011, Dallas. Grants from FAPESP and CNPq}
%Grants or other notes
}

%\titlerunning{Hadrons in AdS/QCD models}        % if too long for running head

\author{W. de Paula \and T. Frederico}

\authorrunning{de Paula and Frederico} % if too long for running head

\institute{W. de Paula and T. Frederico\at
Instituto Tecnol\'ogico de Aeron\'autica, 12.228-900, S\~ao Jos\'e dos Campos, Brazil \\
              \email{wayne@ita.br}
%  \\
}

\date{Received: date / Accepted: date}

\maketitle

\begin{abstract}
We  discuss applications of gauge/gravity duality to describe the
spectrum of light hadrons. We compare two particular 5-dimensional
approaches: a model with an infrared deformed Anti-de Sitter metric
and another one based on a dynamical AdS/QCD framework with
back-reacted geometry in a dilaton/gravity background. The models
break softly the scale invariance in the infrared region and allow
mass gap for the field excitations in the gravity description,
while keeping the conformal property of the metric close to the
four-dimensional boundary. The models provide linear Regge
trajectories for light mesons, associated with specially designed
infrared gravity properties.  We also review the results for
the decay widths of the $f_0'$s into two pions, as overlap integrals
between mesonic string amplitudes, which are in qualitative
agreement with data. 
\end{abstract}
\section{Introduction}
\label{intro}

The AdS/CFT correspondence\cite{Maldacena_Conjecture}  map
$\textit{N}=4$ Super Yang-Mills (SYM) theory in 4D flat space into
Type IIB string field theory in 10D space-time $AdS_{5}\times
S_{5}$. This mapping is such that the strong-coupling regime from
one theory is equivalent to the perturbative regime of the other.
The hope is to develop analytical tools, suitable for the
perturbative regime, to treat the nonperturbative and strongly
interacting region of quantum chromodynamics (QCD).

The holographic principle applied to QCD would allow to describe the
strong interaction and non-perturbative dynamics using gravity
weak-coupling perturbative methods. However, the basic difficulty to
use this method to analyze strong-force physics lies on the fact
that the gauge theory within the AdS/CFT duality is very different
from QCD. In short, $\textit{N}$=4 SYM theory has a conformal
symmetry, whereas QCD breaks this symmetry at low energies and also
the $\textit{N}$=4 SYM theory is supersymmetric, whereas QCD does
not have this symmetry. Consequently, one should modify the AdS
geometry to build a realistic gravity dual of QCD. Applications of
gauge/gravity dualities to ''QCD-like'' gauge theories either start
from specific D-brane setups in ten- (or five-) dimensional
supergravity and derive the corresponding gauge theory properties,
or try to guess a suitable background and to improve it in bottom-up
framework by comparing the predictions to QCD data.

A top-down  approach to seek for a QCD dual starts by considering
D-branes in the theory, which break supersymmetry in part and
introduces flavor. One example is  $N_{f}$ D7 probe branes ($D3-D7$
model\cite{D3D7}), which brings flavor physics to AdS/CFT. The
supergravity side is dual to a four-dimensional $\it{N} = 2$
supersymmetric large-N gauge theory. Even that, a running coupling
constant is found in a type II B $\it{N} = 2$ theory, it still does
not has confinement. There is a vast literature addressing these
issues, see e.g. the review \cite{Erdmenger}. The
failure in getting hadronic spectra suggests that other strategies
in the search for the QCD dual should be envisaged with the guidance
of experimental data to shape, even in a phenomenological way, the
holographic representation of QCD.

Polchinski and Strassler\cite{hw} gave a first step along
these lines. They introduced an infrared (IR) hard-wall in the fifth
dimension, which makes a slice of AdS where a boundary condition at
the QCD scale requires discrete string modes. Following this work, many applications to
study the colorless spectrum from the strong interaction properties
of QCD have been pursued (see e.g. glueballs in \cite{BB03,BB04}
and light hadrons in \cite{Teramond_PRL05}.)
Although, the hard-wall model accounts qualitatively for many
aspects of hadron physics\cite{BB03,BB04,Teramond_PRL05}, it does not lead to linear
Regge trajectories. This difficulty was solved by the introduction
of soft-wall model\cite{Karch} with AdS$_{5}$ geometry and an
additional inert dilaton background field. It leads to linear Regge
trajectories, $m_{n,S}^{2}\sim n+S$, for light-flavor mesons of spin
$S$ and radial excitation $n$. Regge behavior can also be
achieved by IR deformations of the AdS$_{5}$ metric
\cite{Kruczenski,FBT}.

Despite the success of the soft-wall models with AdS$_5$ geometry,
the resulting vacuum expectation value (vev) of the Wilson loop does
not exhibit the characteristic area-law behavior of a confining
theory. The AdS$_5$ metric with the associate conformal invariance
property does not confine by the Wilson loop analysis
\cite{MaldaPRL,Rey01}. Moreover one can raise the issue that the
soft-wall model background with a nontrivial dilaton and AdS$_5$
metric are not a solution of Einstein equations theory. The
analysis of the 5d coupled dilaton-gravity Einstein equations was
performed by Csaki and Reece \cite{Csaki} (see also\cite{Kiritsis})
using the superpotential formalism. They concluded that to solve
those equations with a linear confining background of the dilaton
soft wall model it would require the addition of new degrees of
freedom, like in particular a tachyon. The tachyon-dilaton-graviton
model was pursued in ref. \cite{Batell}.

In order to overcome these difficulties without the introduction of a
tachyon field, in \cite{dePaulaPRD09} was proposed a Dynamical AdS/QCD
model, solution of the dilaton-gravity equations, that leads to a linear
confining background. For the lower excitations, where experimental data
exists an approximate linear trend is found.

The basic amplitudes in the AdS$_5$ deformed inspired models, for
which dynamics provide a whole excitation spectrum, have been
recently promoted to the status of a wave function. This novel
conceptual development concerns the connection between the valence
light-front wave function and the mesonic string amplitude
associated with classical fields propagating in Anti-de Sitter space
\cite{Brodsky_PRL06}. The key point is to realize that the string
mode eigenvalue equation, which carries the breaking of the
conformal symmetry induced either by an infrared boundary condition
or by an effective potential through a coupling of the meson field
to a dilaton and gravity, is analogous to a valence eigenmode
equation. In this way, the holographic realization of the mass
squared operator equation for the valence wave function is achieved.
The free mass squared operator is identified and also an effective
potential which carries the complexity of the coupling of the
valence sector with the full light-front Fock space determined by
the QCD hamiltonian embedding at the same time confinement. The
identification suggested access the content of the so called
iterated resolvent method \cite{pauli} proposed as a way to bridge
the valence physics with the full QCD dynamics in the light-front
Fock space.

We add that the duality between gauge theories and classical
gravity\cite{Maldacena_Conjecture} have been successfully used in
many different contexts. We focus in hadronic physics,
and mention one example that nonconformal holography has also been
applied to describe the elliptic flow in relativistic heavy ion
collisions \cite{QGP}.

In this contribution, we  discuss two particular applications of
gauge/gravity duality to describe the spectrum of light hadrons
\cite{FBT,dePaulaPRD09}. One model has an infrared deformed Anti-de Sitter
metric \cite{FBT} and the another one is based on a dynamical
AdS/QCD framework\cite{dePaulaPRD09}. Both models break softly the scale
invariance in the IR region bringing a gap for the field excitations
in the gravity description, while keeping the conformal
property of the metric at the four-dimensional boundary. As we have
discussed these models provide linear Regge trajectories for
light mesons, associated with specially designed infrared gravity properties.
Furthermore, within the dynamical AdS model
the decay widths of the $f_0'$s into two pions computed as overlap
integrals between mesonic string amplitudes \cite{dePaula:2009za} is
found in qualitative agreement with data \cite{pdg}.
\section{Infrared deformed AdS/QCD model}
In the following, we discuss the holographic calculation of hadron
spectrum from two different perspectives \cite{FBT} and
\cite{dePaulaPRD09,dePaula:2009za}. The basic idea is that a hadron
can be described by classical fields propagating in the
AdS$_5$ gravity background with IR deformation, where the metric is
written as
\begin{equation}
ds^2 = e^{-2A(z)}\left(\eta_{\mu\nu}dx^{\mu}dx^{\nu}-dz^{2}\right)\label{AdS_metric_warp},
\end{equation}
where  $\eta_{\mu\nu}$ = $\left(1,-1,-1,-1\right)$, $x^{\mu}$ =
$\{x^{1}, x^{2}, x^{3}, x^{4}\}$ and $z$ is the holographic coordinate.

In holographic dual models the propagating states corresponding to
hadrons in the 4-dimensional boundary, are the normalizable modes of
5D fields. The  gauge/gravity dictionary identifies the eigenvalues
with the mass squared of the hadron in the boundary gauge theory.
The mode equation written in a
Sturm-Liouville form for the reduced amplitudes, $\psi _{n}$,
appears generally as
\begin{equation}
\left[ -\partial _{z}^{2}+\mathcal{V}(z)\right] \psi
_{n}=m_{n}^{2}\psi _{n}\ , \label{SL_AdS}
\end{equation}
where $m_{n}$ is the hadron mass. For the particular case of the AdS
metric $A(z)=\log (z)$ the eigenvalue problem has a continuum
spectra. For mesons dual  to modes of a scalar $(S)$ and vector $(V)$
fields, and baryons $(B)$  dual to a spin 1/2  fields, the effective
potentials are  respectively written as:
\begin{eqnarray}
\mathcal{V}_{S}(z)=\frac{1}{z^2}\left(\frac{15}{4}+m_{5}^2\right),
~~ \mathcal{V}_{V}(z)=\frac{1}{z^2}\left(\frac{3}{4}+m_{5}^2\right),
~~\mathcal{V}_{B}(z)=\frac{m_{5}}{z^2}\left(m_{5}\mp1\right),
\label{potv}
\end{eqnarray} where the sign $\pm$ depends on the fermion/baryon chirality
$i\gamma^{5}\Psi_{\pm}=\pm\Psi_{\pm}$. Within the AdS/CFT
correspondence the full mode amplitude should behave as $z^{\tau}$,
where $\tau = \Delta - \sigma$ (conformal dimension minus spin) is
the twist dimension for the corresponding interpolating operator
that creates the given state configuration \cite{hw}. The
five-dimensional mass is chosen to fit the twist dimension as i)
$m_{5}=\sqrt{\tau(\tau-4)}$ for scalars, ii)
$m_{5}=\sqrt{\tau(\tau-4)+3}$ for vector and iii) $m_{5}=\tau-2$ for
baryons. The outcome of the potentials (\ref{potv}) is a continuum spectrum.

The discrete spectrum solution of (\ref{SL_AdS}) demands a change of
the AdS$_{5}$ effective potentials given by (\ref{potv}).
The hard-wall model\cite{hw} results in contrast to the observed
linear Regge behavior of the squared masses\cite{Anisovich00} with
a quadratic dependence on $n$ \cite{Teramond_PRL05}. The dilaton
with AdS$_5$ background successful for mesons \cite{Karch}, was
unable to confine fermionic modes. Then, in \cite{FBT} it was
suggested to perform the substitution of the twist by
$\tau+\lambda^2 z^2$ in the effective potential (\ref{potv}) of the
Sturm-Liouville equation, accounting for the almost universal slope
of the light hadron trajectories \cite{Anisovich00,jmrichard}. In
particular, the 5D mass of the hadronic mode becomes proportional to
$\lambda^2z^2$. Recent approaches that treats fermionic
modes indeed profits from changing  the mass with $z^2$
\cite{Schmidtm2}.

After the introduction of the soft IR modification of the effective
mode potential (\ref{potv}) for baryons and mesons $(M)$ according
to the prescription suggested by \cite{FBT} is
\begin{equation}
\mathcal{V}_{B}(z)=\frac{1}{z^2}\left((L+1)(L+1\mp1)+[2(L+1)\pm1]\lambda^2~z^2+\lambda^4~z^4\right),~~
\mathcal{V}_{M}(z)=\frac{1}{z^2}\left((\lambda^2~
z^2+L)^2-\frac{1}{4}\right),\label{SL_LTM}
\end{equation}
%\begin{eqnarray}
%\mathcal{V}_{B}(z)=\frac{1}{z^2}\left((L+1)(L+1\mp1)+[2(L+1)\pm1]\lambda^2~z^2+\lambda^4~z^4\right) \ , \label{SL_LTB}\\
%\mathcal{V}_{M}(z)=\frac{1}{z^2}\left((\lambda^2~
%z^2+L)^2-\frac{1}{4}\right),\label{SL_LTM}
%\end{eqnarray}
with $L$ being naively associated with the orbital angular momentum
of the internal hadron state, which is introduced by accounting the
dimension of the number of covariant derivatives introduced in the
interpolating operator \cite{Brodsky_PRL06}. The
light meson  and baryon spectra is given by:
\begin{equation}
M_{M}^{2}=4\lambda^2~\left(N+L+\frac{1}{2}\right)\; , \,
M_{B}^{2}=4\lambda^2~\left(N+L+\frac{3}{2}\right),
\end{equation}
exhibiting the linear Regge behavior. Experimental data \cite{pdg}
just gives $\lambda\sim 1$~GeV. Both formulas are quite appropriate
to describe the light hadron spectrum, in particular a link between
the slope and intercept is established also in the baryonic case.
The delta spectrum is quite well described \cite{FBT}, while the
nucleon demands a different relation between the slope and
intercepted \cite{forkelklempt} (see also \cite{jmrichard,Brodsky_Spectrum}.)

The asymptotically AdS$_5$ metrics which give linear Regge trajectories
depends on the quantum number $L$, suggesting that the
5D description could be not enough to permit modes in an universal
metric background independent of the hadron. In this direction
fluctuations of the background in a 10-dimensional supergravity
could be much more appealing, however it is still missing a mass gap
\cite{dePaulaJHEP}. The effective potential of the Sturm-Liouville
equation for a deformed AdS reads:
\begin{equation}
\mathcal{V}_{B}(z)=m_{5}\; e^{-A}\left(m_{5}e^{-A}-A'\right) \ ,
\label{vdads}
\end{equation}
with $m_{5}=\tau-2$ and the confining effective potential
(\ref{SL_LTM}) is delivered through (\ref{vdads}) by an analytical
form of the metric given by:
\begin{equation}
A=-\log \left(\frac{1}{\lambda~z}+\frac{\lambda~z}{L+1}\right).
\end{equation}
Note that the metric has the quantum number $L$ dependence. For
mesons the correspondent form of the metric is so far found by
numerical means and the details were discussed in \cite{FBT}.
\section{Dynamical AdS/QCD model}
The bottom-up proposal of a dynamical AdS/QCD model with a
dilaton-gravity back-reacted geometry \cite{dePaulaPRD09}, came to
address the issue raised by the vev of
the Wilson loop in non-deformed AdS$_5$ metric, which does not
exhibit the characteristic area-law behavior that a linearly
confining static quark-antiquark potential would generate. A
second, common shortcoming of a soft-wall background is that it is
not solution of the 5D Einstein equations (see e.g.
\cite{Kiritsis}).  Within our proposal of a back-reacted geometry of
the dilaton-gravity model the Regge trajectory of the mesonic mass
spectrum presents an approximate linear behavior. The action for
five-dimensional gravity coupled to a dilaton field is:
\begin{eqnarray}
S = \frac{1}{2k^{2}}\int d^{5}x \sqrt{g} \left( -\emph{R} -
V(\Phi)+\frac{1}{2}g^{MN}\partial_{M}\Phi\partial_{N}\Phi\right),
\label{actiongd}
\end{eqnarray}
where the metric is given by (\ref{AdS_metric_warp}), $k$ is the Newton constant and $V(\Phi)$ is
the scalar field potential. The equations of motion are a coupled set of Einstein equations with
solutions satisfying the following relations
\begin{equation}
\Phi'= \sqrt{3 A'^{2} + 3 A''}~,~V(\Phi) = \frac{3 e^{2A}}{2}
\left(A''-3 A'^{2}\right), \label{constrain}
\end{equation}
which completely determines the dilation properties. This is the strategy of
\cite{dePaulaPRD09} to built the dilaton-gravity background with the
metric satisfying the scale invariance in the UV region and the
criteria of Wilson loop area-law in the IR region, which says that
$A~z^t$ ($t\geq 1$)~\cite{Kiritsis}. These constraints are the guiding
physics that the phenomenologically parameterized metric should
satisfy.

{\it Mesons with spin.} The 5D action for a gauge field
$\phi_{M_{1}\dots M_{S}}$ of spin $S$ in the dilaton-gravity
background is given by~\cite{Karch}
\begin{equation}
I = \frac{1}{2} \int d^{5}x \sqrt{g} e^{-\Phi}\left(\nabla_{N}
\phi_{M_{1}\dots M_{S}} \nabla^{N} \phi^{M_{1}\dots M_{S}} \right).
\end{equation}
As in \cite{Karch} and \cite{KatzLewandowski} the axial gauge is
used and new spin fields $\widetilde{\phi}_{\dots}=
e^{2(S-1)A}\phi_{\dots}$ introduced. The substitution
$\tilde{\phi_{n}}= e^{B/2}\psi_{n}$ in the mode equation gives it in
the Sturm-Liouville form of eq. (\ref{SL_AdS}), with~${\mathcal
V}_M(z)=\frac{B'^{2}(z)}{4} -\frac{B''(z)}{2}$, where $B = A (2S-1)
+ \Phi$. The choice of the warp factor
\begin{equation}
A(z)=\log (z \Lambda_{QCD})+ \frac{1+\sqrt{3}}{2S+\sqrt{3}-1}\frac{(z\Lambda _{QCD})^{2}}{%
1+e^{(1-z\Lambda _{QCD})}},  \label{cnew}
\end{equation}%
reproduces the linear Regge trajectories \cite{dePaulaPRD09}, with
the associated dilaton field and potential obtained from eq.s
(\ref{constrain}). An analytical approximation to the spectrum for $%
\Lambda _{QCD}=0.3$ GeV is (in units of GeV)
\begin{equation}
m_{n,S}^{2}\simeq \frac{1}{10}\left( 11n+9S+2\right) ,~~~~~\left( n\geq 1\right)
\end{equation}%
which implements the approximate universality of the linear
trajectory slopes for light flavors explicit.

{\it Scalar and pseudoscalar mesons.} We write the metric as
\begin{equation}
A(z)= \log(z \Lambda_{QCD}) + \frac{(\xi z\Lambda _{QCD})^{2}}{%
1+e^{(1-\xi z\Lambda _{QCD})}},  \label{cscalar}
\end{equation}%
to describe the $f_{0}$ and pion families with a single parameter
$\xi$. For the $f_{0}$ family we found $\xi=0.58$
\cite{dePaula:2009za}. Comparing to the vector sector, the slope of
the Regge trajectory for the scalar excitations is reduced. This
means that the size of the scalar ground state $f_0(600)$ should be
larger than the size of other light mesons ground states, see the
left-frame of fig. 1, where we compared the  $f_0(600)$ and $\pi$
reduced amplitudes. We mention that, scalar mesons were also
analyzed in (\cite{Schmidt_Scalar} and \cite{Colangelo_Scalar}).

The pion Regge trajectory has an experimental slope of about 1
GeV$^2$ and near twice the value found for the $f_0$ family.
Therefore, the scaling factor of the holographic coordinate for
pseudoscalars is larger than the one corresponding to the $f_0'$s.
Indeed by equating IR effective potentials of the pion and  higher
spin meson $\xi=0.76$ is found, and a value of $\xi=0.88$ allows a
fine-tuning of the pion mass trajectory. The small pion mass is
implemented by changing the 5D mass as $m_{5}^{2} \rightarrow
m_{5}^{2}-\overline\lambda z^{2}$ with
$\overline\lambda=2.19$GeV$^2$\cite{dePaulaIJMPD}. The $f_0$'s partial
decay width into $\pi\pi$ can be obtained from
the overlap integral of the normalized string amplitudes
(Sturm-Liouville form) in the holographic coordinate dual to the
scalars $(\psi_n)$ and pion $(\psi_\pi)$ states,
\begin{equation}
h_{n} =\overline\lambda\,\Lambda_{QCD}^{-\frac32}\int^\infty_0 dz
~\psi_\pi^2(z)\psi_n(z), \label{dwover}
\end{equation}
where the parameter $\overline\lambda$ gives the natural strength
of the transition amplitude as it is the scale for the coupling
between the pion and a scalar field, as has been obtained through
the pion mass shift. The factor of $\Lambda_{QCD}$ is introduced to
provide the correct dimension of the decay width. We find that
$\lambda\, \Lambda_{QCD}^{-\frac32}= 13~[$GeV$]^\frac12$, for
$\Lambda_{QCD}=$ 0.3 GeV, giving the results shown in the left-frame
of fig. 1 compared to the experimental data. First, $f_0(980)$ has a
mixing angle of 20$^0$\cite{Bediaga} to fit the partial width. Furthermore, we
observe that $f_0\to\pi\pi$ decay width decrease fast as radial
excitation of the scalar increase. The overlap between the the pion
wave functions and the excited $f_0$ modes  with  nodes (see
left-frame of fig.1) explains such a decrease\cite{dePaulaNPBPS}.

\begin{figure}
\centering
\epsfig{file=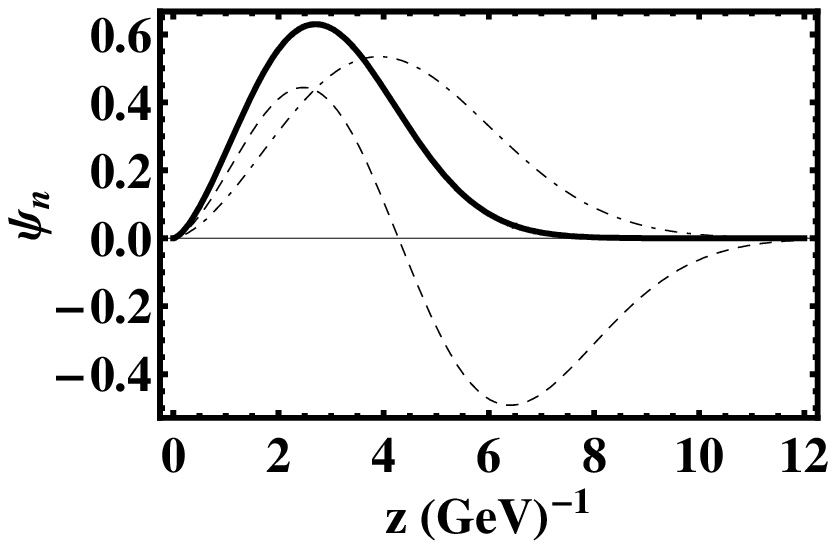,width=6.cm,angle=0}
\epsfig{file=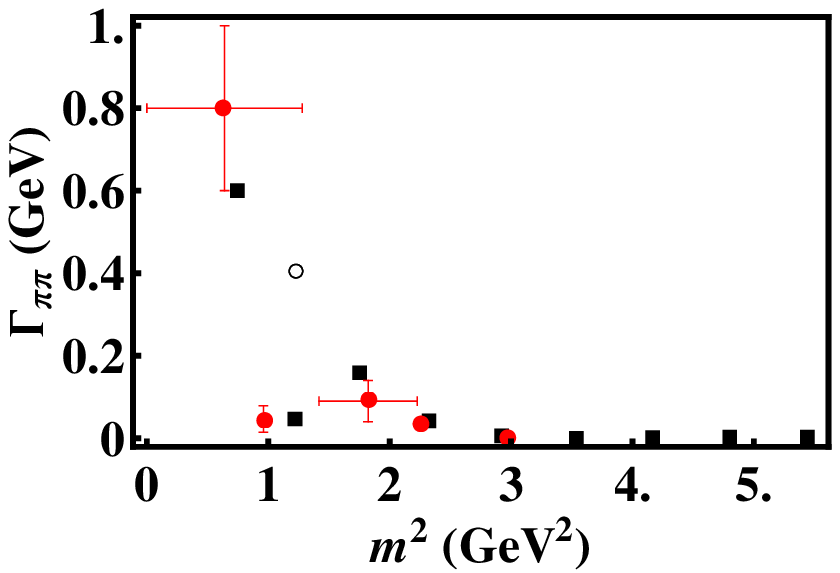,width=5.7 cm,angle=0}
\caption{Results for the dynamical AdS model. {\it Left frame:}
reduced amplitudes for the pion (solid line), sigma (dot-dashed
line) and $f_0(980)$ (dashed line). {\it Right frame:} $f_0\to
\pi\pi$ decay partial width v.s. $f_0$ mass squared compared to data
\cite{pdg}. The calculations are given by full boxes. The empty
circle corresponds to the calculation of $f_0(980)\to\pi\pi$ without
considering the mixing angle of 20$^0$ \cite{dePaula:2009za}.}
\label{fig:1}       % Give a unique label
\end{figure}
\section{Summary}
We discussed some aspects of the bottom-up strategy to reveal
aspects of the duality between gravity and QCD, by
phenomenological studies of the light-hadron spectrum and decay of
scalar mesons. Two particular 5D approaches were addressed. One
model is built based on an infrared deformed Anti-de Sitter metric
\cite{FBT}. The other model has a dynamical AdS/QCD framework with
back-reacted geometry in a dilaton/gravity background
\cite{dePaulaPRD09}. We stress that the methods used for the
dynamical AdS/QCD model, solution of the five-dimensional
Einstein-dilaton equations applies to essentially all asymptotically
AdS$_{5}$ with a Poincar\'{e}-invariant boundary.
Both holographic models break the scale invariance in the infrared
region and allow to describe the light-meson spectrum with a gravity
description. They keep the conformal property of the
metric close to the four-dimensional boundary. Linear Regge
trajectories for light mesons are associated with specially designed
infrared metric properties. The results obtained for the decay
widths of the $f_0\to\pi\pi$s, as overlap integrals between mesonic
string amplitudes, compared to the data seems encouraging. A
challenge remains for the description of baryons within the
dynamical AdS/QCD framework, which decouples from the dilaton field
and for them it is enough a polynomial metric, which breaks the
area-law criteria.

\end{document}